\documentclass[conference]{IEEEtran}
\usepackage{blindtext, graphicx}
\usepackage{color}
\usepackage{cite}
\usepackage{amsmath,amsfonts,amsthm,bm} 
\newtheorem{theorem}{Theorem}
\usepackage{siunitx}
\usepackage[normalem]{ulem}
\usepackage{epstopdf}

\IEEEoverridecommandlockouts

\usepackage{comment}

\usepackage{graphicx} 
\usepackage{float}
\usepackage[table]{xcolor}
\usepackage{soul}
\usepackage{amsmath}
\usepackage{amsthm}
\usepackage{amsfonts}
\usepackage{subcaption}
\usepackage[numbers]{natbib}
\usepackage{blkarray}
\usepackage{algorithm}
\usepackage[noend]{algpseudocode}
\usepackage{multirow}
\usepackage{mathtools}
\usepackage{array}
\usepackage{blkarray}
\usepackage[hyphens]{url}

\usepackage{adjustbox}
\usepackage{lipsum}

\newcolumntype{C}[1]{>{\centering\arraybackslash}p{#1}}

\title{
Physics-Informed Building Occupancy Detection: a Switching Process with Markov Regime}

\author{
Amir-Mohammad Esmaieeli-Sikaroudi$^1$, Boris Goikhman$^2$, Dmitri Chubarov$^2$,  Hung Dinh Nguyen$^3$, \\  Michael Chertkov$^1$, and Petr~Vorobev$^3$\\
\IEEEauthorblockA{$^1$University of Arizona, Tucson, AZ.
Emails: \{amesmaieeli, chertkov\}@arizona.edu
}
\IEEEauthorblockA{$^2$Airvoice, R\&D Center, Wilmington, DE. Emails: \{bg, dc\}@airvoice.global
}
\IEEEauthorblockA{$^3$Nanyang Technological University, School of EEE, Singapore. Emails: \{hunghtd, petr.vorobev\}@ntu.edu.sg
}
}

\date{}

\begin{document}

\maketitle

\begin{abstract}
Energy efficiency of buildings is considered to be one of the major means of achieving the net-zero carbon goal around the world. The big part of the energy savings are supposed to be coming from optimizing the operation of the building heating, ventilation, and air conditioning (HVAC) systems. There is a natural trade-off between the energy efficiency and the indoor comfort level, and finding an optimal operating schedule/regime requires knowing the occupancy of different spaces inside of the building. Moreover, the COVID-19 pandemic has also revealed the need to sustain the high quality of the indoor air in order to reduce the risk of spread of infection. Occupancy detection from indoor sensors is thus an important practical problem. In the present paper, we propose detection of occupancy based on the carbon dioxide measurements inside the building. In particular, a new approach based on the, so-called, switching auto-regressive process with Markov regime is presented and justified by the physical model of the carbon dioxide concentration dynamics. We demonstrate the efficiency of the method compared to simple Hidden Markov approaches on simulated and real-life data. We also show that the model is flexible and can be generalized to account for different ventilation regimes, simultaneously detecting the occupancy and the ventilation rate.
\end{abstract}

\section{Introduction}

Energy efficiency is considered to be the largest measure to stop the energy demand growth within the net-zero goal according to the International Energy Agency report \cite{IEA_efficiency}. Building energy efficiency, in particular, is a promising solution, since buildings are responsible for around $30\%$ of the total energy consumption \cite{economidou2020review}. Heating, ventilation, and air conditioning systems (HVAC) have traditionally contributed to about half of the building's total energy consumption. HVAC systems typically work according to some standard pre-specified schedule that does not explicitly take into account the indoor air parameters (apart from temperature). It is not surprising, that optimization of HVAC operation is thought to be one of the major contributors to energy savings. Moreover, due to rather slow rate of thermal processes, controlling HVAC systems provides an excellent opportunity for grid services, like demand-response \cite{koch2011modeling}. 

Since HVAC systems are responsible for the indoor air quality, there exists a natural trade-off between the comfort level and energy efficiency. Moreover, after the outbreak of the COVID-19 pandemic, it became clear that the indoor air condition is an important factor that can contribute to the rate of the disease spread. Various operation guidelines for HVAC systems in regards to COVID-19 pandemic were widely discussed \cite{guo2021review,ding2020hvac}. Moreover, there is a growing body of research that suggests that even moderate air quality degradation can lead to a visible reduction in people's cognitive skills \cite{allen2016associations}. It is reasonable to assume that the problem of HVAC scheduling/control should explicitly take into account the actual state of the building occupancy. 

The problem of estimation of building occupancy (i.e., the number of people in each room) has been actively discussed in the research literature for more than a decade. Numerous data-driven methods were suggested, more details can be found, for example, in reviews \cite{chen2018building,rueda2020comprehensive}. One of the main practical issues with many machine learning methods is the need to have comprehensive labeled data sets to train the models. This can be difficult to obtain because one needs to count the number of occupants in every room over an extended period of time. One of the efficient methods that  was proposed for occupancy detection that can avoid this difficulty is the use of the Hidden Markov Models (HMM) where the room's occupancy is represented by the hidden states and different indoor air parameters are the observable variables \cite{candanedo2017methodology}. Of particular interest is the use of more advanced models, such as Auto-Regressive Hidden Markov Models (ARHMM) \cite{ai2014occupancy} that can potentially take into account correlation between indoor air parameters (e.g., carbon dioxide concentration) at different moments of time. Such an approach looks reasonable since the physical dynamics of air properties include their gradual dynamics over time, rather than abrupt change with the changes in the occupancy.

In the present manuscript, we have developed a \emph{physics informed} approach for occupancy detection using the carbon dioxide measurements. Starting from the fundamental dynamic equations, we show that the problem naturally corresponds to switching auto-regressive process with Markov regime \cite{ephraim2002hidden}. We link the occupancy and the ventilation rate to the parameters of the model, and demonstrate the model performance on the simulated data sets and real-life measurements.

\section{Occupancy Estimation: a Hidden-Markov Model-based approach}

The presence of occupants in the building leads to variation in the indoor environment parameters. The most notable parameters that are affected are the carbon dioxide concentration, temperature, and humidity, with carbon dioxide being the most sensitive one. Carbon dioxide is produced almost exclusively by people in the building with its concentration being rather stable outside. In the following, we will use only the carbon dioxide measurements for occupancy estimation, but our models can be rather easily generalized to include more variables. 

As discussed in the Introduction, the Hidden Markov Model (HMM) approach is a suitable method for occupancy detection since it can be developed without the need for labeled training data. HMM represent a powerful inference tool and they have been widely used in different areas of science and engineering for several decades. A model is represented by a set of 'hidden' states that can not be directly observed and the measurements that are somehow related to the hidden states. Both hidden states and measurements represent stochastic processes, therefore, the solution to the estimation problem is naturally probabilistic. 

Let us assume that the HMM can be represented by $N$ hidden states, and denote the state at time $t$ as $S_t$. For the occupancy problem, a natural choice of states would be the number of occupants; however, we can also have additional states that correspond, for example, to different operating regimes of the ventilation system. We will denote the system observables at time $t$ as $y_t$ - this could be a scalar or a vector in the case of multiple measurements in each moment of time. We also denote the set of the system parameters as $\theta$ - these can include the transition probabilities between the states, probabilities of different observations while at a certain state, level of noise, etc. Whenever possible, we would like to refer the model parameters to the properties of the underlying physical system. In regards to the occupancy estimation task, we need to address the following two problems:

\emph{Problem 1}. Assuming some values for the model parameters $\theta$, what would be the most probable sequence of the hidden states under the condition of the present sequence of observables? We will need to maximize the following conditional probability $P(y_1,y_2,...,y_T|\theta)$ with respect to the set of variables $S_t$. 

\emph{Problem 2}. In the case we are uncertain about the model parameters, what would be their values that maximize the probability of observing the measured sequence? To achieve this, we will need to maximize the probability $P(y_1,y_2,...,y_T|\theta)$ in respect to the parameters $\theta$.

For more detailed discussion of the stated problems, we refer to an excellent tutorial paper \cite{rabiner1989tutorial}. One important note should be made here: when maximizing the probabilities for either of the above stated problems, it is possible to \emph{overfit} the model - i.e. find the solution that is formally more probable than the true one, or arrive at a local minimum, that will be far away from the true system parameters and states. It is possible to minimize the chance of both effects by properly choosing the model type and the initial values for the model parameters - using the properties of the underlying physical model.

\subsection{Model Formulation}

The straightforward approach to formulating a HMM model for occupancy detection is to assume that the hidden states correspond to the room's occupancy and the observables correspond to the measurements of the carbon dioxide concentration. In such a model, the carbon dioxide concentration in a room at some moment of time is assumed to be dependent only on the occupancy at the same moment of time, and it does not depend on the previous values of both occupancy and carbon dioxide concentration. While such a model is very simple to deal with, since the underlying physical system does not directly correspond to such a model, it can have rather poor accuracy. In fact, the concentration of carbon dioxide has certain dynamics, and generally its present value depends on its values and the values of the occupancy at all the previous moments of time. In \cite{ai2014occupancy} a so-called, auto-regressive Hidden Markov Model was proposed to address the occupancy detection problem. However, the model was stated in a rather general mathematical form, but no physical justification or derivation was provided. In the present work we are proposing a \emph{physics-based} HMM-type model, namely, \emph{switching auto-regressive process with Markov regime} \cite{ephraim2002hidden}, which we derive from the first physical principles. Such an approach will provide the model with the least number of parameters - that is beneficial for preventing overfitting, and additionally, it will allow us to choose reasonable starting values for the unknown model parameters.

We start our derivation of the auto-regressive Markov switching model by stating the dynamic equation for a carbon dioxide concentration in a room. We assume that carbon dioxide is produced by people present in the room and is removed by the ventilation system. The basic equation can be written in the following form \cite{cali2015co2}:
\begin{equation}\label{eq:main_CO2}
\frac{d x(t)}{dt} = -\frac{1}{\tau}(x(t) - x^{(0)}) + n(t) r
\end{equation}
Here $x(t)$ denotes the concentration of the carbon dioxide in the room as a function of time, $x^{(0)}$ is the ambient concentration of carbon dioxide, $n(t)$ - is the number of people present in the room as a function of time, $r$ - is the rate of carbon dioxide production by a single person, and $\tau$ - is the effective ventilation time. The first term in the right-hand side of \eqref{eq:main_CO2} represents the effect of ventilation while the second term is due to carbon dioxide produced by people present in the room. 

It is convenient to switch to a new variable $y(t) = x(t)  - x^{(0)}$, then equation \eqref{eq:main_CO2} can be written in a more compact form:  
\begin{equation}\label{eq:simple_CO2}
\dot{y} = -\frac{1}{\tau} y + n(t) r
\end{equation}

We are now ready to state the following Theorem:

\begin{theorem}\label{theorem:AR-HMM}
Assume carbon dioxide dynamics obeys equation \eqref{eq:simple_CO2}. Then, the problem of room occupancy estimation can be represented as a \bf{switching auto-regressive process of order one (and a drift) with Markov regime}. 
\begin{proof}
Let us first rewrite equation \eqref{eq:simple_CO2} in discrete time. For this, we assume that the room occupancy and the ventilation rate remain constant during each time-step (which is a realistic assumption) and integrate both sides of the equation from the time instant $t$ to $t+1$ (assuming the time step size is $\Delta t$). We have:
\begin{equation}\label{eq:CO2}
y_{t+1} = e^\frac{-\Delta t}{\tau} y_t + (1-e^\frac{-\Delta t}{\tau})r n_t 
\end{equation}
Introducing noise (which can be thought of as a fluctuation in carbon dioxide production rate by people) and making the appropriate change in denotations, this relation can be written in the following way:
\begin{equation}\label{eq:AR}
y_{t} = c_{S_t} y_{t-1} + \mu_{S_t} +  W_{S_t}
\end{equation}
Here $S_t$ denotes the state of the Hidden Markov Model at time $t$, the auto-regression coefficient at time $t$ is $c_{S_t}=\exp(-\Delta t/\tau)$, the 'drift' term is $\mu_{S_t} = (1-\exp(-\Delta t/\tau)r n(t-1)$ which is dependent on the room occupancy. The last term, $W_{S_t}$ is the noise term while at state $S_t$. The model \eqref{eq:AR} represents the switching auto-regressive process, thus our theorem is proved.
\end{proof}
\end{theorem}

We note, that our model can describe the system that has multiple ventilation regimes. In this case, our model will have more hidden states, each corresponding to one value of occupancy and a certain ventilation rate. It is convenient, that the auto-regression coefficient $c_{S_t}$ is only dependent on the ventilation rate, while the drift term $\mu_{S_t}$ depends on the occupancy. In the following, we will denote $c$ and $\mu$ terms corresponding to state $S_t=i$ simply as $c(i)$ and $\mu(i)$ respectively. 

We note that representation of the occupancy problem as an auto-regression HMM model (without a drift term) was already discussed before in \cite{ai2014occupancy} as a general formulation. However, our approach links the model parameters to the real physical properties of the system. Compared to representation in \cite{ai2014occupancy} our model also contains the 'drift' term, which is specifically determined by the occupancy level. 

\subsection{Model analysis}

Model \eqref{eq:AR} represents the first-order auto-regressive model with a drift, where both auto-regression coefficient and the drift are dependent on the system's current hidden state. In order to give more intuition on how the model can be used to determine the coefficients $c_{S_t}$ and $\mu_{S_t}$ let us first analyze a period when the model stays in a single hidden state $S_t=i$ over $T$ time steps ($S_t=i, t=1,2,..T$). For this period, equations \eqref{eq:AR} reduce to a classical AR model (with a drift), and the coefficients can be estimated using Yule-Walker equations \cite{shumway2000time}:
\begin{equation}\label{eq:c_hat}
    \hat{c}_{i} = \frac{\sum\limits_{1}^{T-1}\left(y_t - \overline{{y}_t} \right)\left(y_{t-1}-\overline{{y}_{t-1}}\right)}{\sum\limits_1^{T-1}\left(y_t - \overline{{y}_{t-1}} \right)^2}
\end{equation}
\begin{equation}\label{eq:mu_hat}
    \hat{\mu}_{i} = \overline{{y}_{t}} - \hat{c}_{i}   \overline{{y}_{t-1}} 
\end{equation}
where we have made the following denotations:
\begin{equation}\label{eq:sigmas}
    \overline{{y}_t} = \sum\limits_{1}^{T}y_t; \qquad \overline{{y}_{t-1}} = \sum\limits_{1}^{T} y_{t-1}
\end{equation}

Generalization of these equations to a Markov model with multiple states is rather straightforward and can be achieved by multiplying the initial equation \eqref{eq:AR} by the probability of observing state $i$ at time $t$ conditioned on the observed sequence of $y$:

\begin{equation}
    q_t(i) = P(S_t = i | Y, \theta ) 
\end{equation}
here $Y = \{y_1,y_2,...,y_T\}$ denotes the sequence of observations, and $\theta$ - the set of model parameters under which the probability is assessed. 

One can show that the new values for the estimated coefficients $\hat{c_i}$ and $\hat{\mu_i}$ are given by the following equation: 

\begin{equation}\label{eq:coeffs_M}
    \left[\hat{c}_{i}\;\hat{\mu}_{i}\right] = -D_i^T E_i^{-1}
\end{equation}
where the vector $D_i$ and matrix $E_i$ are given by the following relations: 
\begin{equation}\label{eq:D_vec}
    D_i = \left[\sum\limits_{1}^{T} q_t(i) y_t y_{t-1}\,\, \sum\limits_{1}^{T} q_t(i) y_t\right]^T
\end{equation}

\begin{equation}\label{eq:E_mat}
    E_i = \begin{bmatrix}
 \sum\limits_{1}^{T} q_t(i) y_{t-1}^2 & \sum\limits_{1}^{T} q_t(i) y_{t-1} \\
 \sum\limits_{1}^{T} q_t(i) y_{t-1} & \sum\limits_{1}^{T} q_t(i) \\
    \end{bmatrix} 
\end{equation}

Expressions \eqref{eq:coeffs_M}-\eqref{eq:E_mat} allow to iteratively update the values of $\hat{c_i}$ and $\hat{\mu_i}$ provided the probabilities $q_t(i)$ are estimated from the parameter values at the previous iteration. This is essentially a realization of the Baum-Welch algorithm \cite{rabiner1989tutorial} for our model. In the present manuscript we will use another method for parameter update, namely, segmental $k$-means algorithm, also known as Baum-Viterbi algorithm \cite{ephraim2002hidden}.

\section{Inference}

Let us define the likelihood for the sequence of observed variables by $\mathcal{L}$. We assume that $\Lambda$ refers to the state transition matrix, and the likelihood of transitions controls how often a state transition can occur.

\begin{equation}\label{eq:likelihood}
\mathcal{L}=\sum_t log(n(Y_t,y_t,\sigma))+\sum_t log(\Lambda(S_t,S_{t-1}))
\end{equation}
 In equation \eqref{eq:likelihood}, the likelihood depends on the latent variable for the number of occupants and the ventilation time ($n(t)$ and $\tau$ from equation \eqref{eq:CO2}). 
 We employ the Expectation Maximization (EM) to obtain both occupancy sequence and ventilation time. EM is an iterative algorithm used to estimate the parameters of statistical models that maximize the probability of the observed variables, particularly when the data is incomplete or has hidden variables. It operates in two main steps: the Expectation (E) step and the Maximization (M) step. In the E step, given an initial guess of the model parameters, the algorithm estimates the most likely sequence of the hidden variables based on the observed data. In the M step, the algorithm maximizes the likelihood function by updating the model parameters using the state sequence data from the E step. These steps repeat alternately until convergence, where the parameter estimates stabilize \cite{do2008expectation}. In the E-step we use Viterbi algorithm \cite{rabiner1989tutorial} to obtain the most likely occupancy sequence, and in M-step we use Maximum Likelihood to update the parameters. This approach was proposed as Expectation-Maximization Viterbi Algorithm in the literature and its convergence properties were investigated in \cite{nguyen2005expectation}.

 The Viterbi algorithm is a dynamic programming technique used for finding the most likely sequence of hidden states in a HMM, given a sequence of observed events. In HMMs, states are not directly visible, but the output or observations dependent on these hidden states are. The algorithm works by calculating the maximum probability of reaching each state at a given time step, based on previous probabilities and the transition model. There are various optimized implementations of Viterbi algorithm that efficiently prune less likely paths, focusing only on the most probable paths as it progresses through the observation sequence, ensuring optimal results without computing every possible state sequence \cite{viterbi1967error,churbanov2008implementing}.

\section{Occupancy Estimation Results}

In this section, we present the validation of our developed method/algorithms for occupancy estimation. We test the method on two types of datasets: synthetic data, where the carbon dioxide concentration was simulated using equation \eqref{eq:main_CO2}, and the real-life experimental data set. The latter was obtained from measuring the carbon dioxide concentration in one of the small conference rooms at the EEE School at Nanyang Technological University. The room occupancy was tracked varying from $0$ (non-occupied room) to the occupancy of $4$ people. The carbon dioxide concentration measurements have a resolution of one minute.   

\begin{figure}[h!]
    \captionsetup{justification=centering}
    \centering
    \includegraphics[scale=0.55]{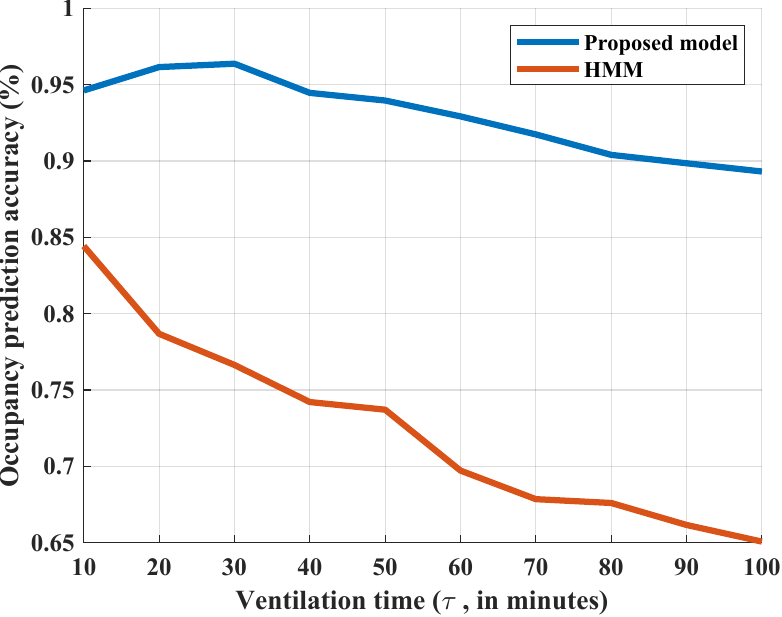}
    \caption{Performance of our proposed and simple HMM model for different ventilation rates. As the ventilation rate gets smaller (i.e., larger values of $\tau$) the performance of HMM model greatly deteriorates. }
    \label{fig:HL_HMM_ARHMM}
\end{figure}

Before proceeding to the actual datasets, we provide a comparison of the occupancy prediction accuracy between our proposed model and a simple "classical" Hidden-Markov Model. We generate a synthetic dataset with different values of the ventilation rate (determined by the ventilation time $\tau$) and compare the performance of both models. Figure \ref{fig:HL_HMM_ARHMM} shows the performance of both models for varying ventilation rates. We see that as the ventilation time $\tau$ gets bigger (i.e., corresponding to slower carbon dioxide dynamics), the simple HMM model loses the accuracy. This is the result of the fact that the carbon dioxide concentration follows the occupancy with a bigger delay, which the model completely ignores. On the other hand, our proposed model shows consistently high accuracy, which only slightly falls with the big increase in ventilation time. We note that the ventilation time $\tau=100$ minutes or greater that corresponds to the right part of the Figure \ref{fig:HL_HMM_ARHMM} is quite realistic for many rooms/buildings.

\begin{figure}[H]
    \captionsetup{justification=centering}
    \centering
    \includegraphics[scale=0.5]{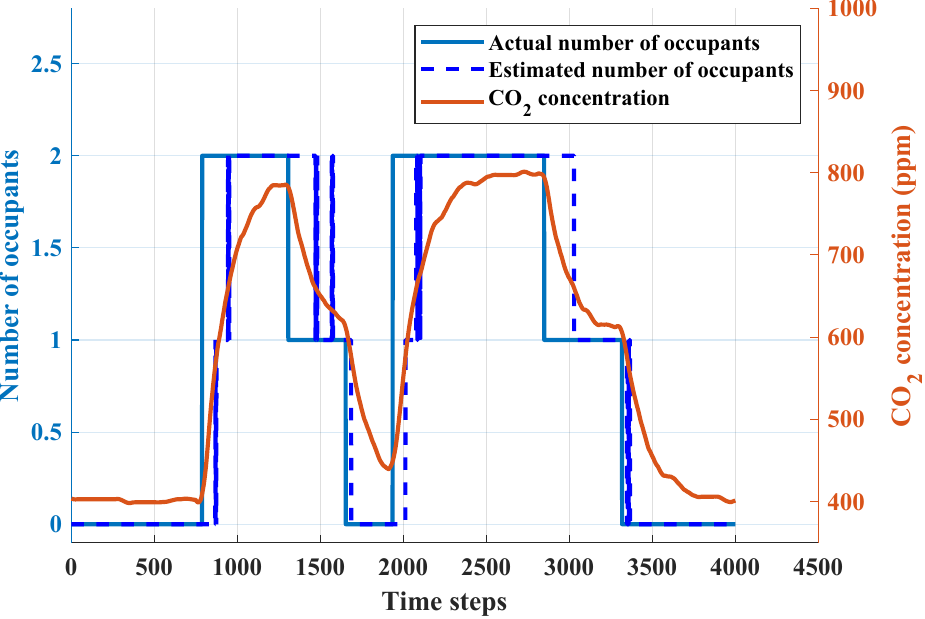}
    \caption{Performance of the simple HMM model on a synthetic dataset. It is visible how the model often has significant delays in correct prediction of the occupancy. The overall accuracy is around $69.78\%$. }
    \label{fig:HMM_SYN}
\end{figure}

Let us now move to check our model performance in regards to occupancy prediction accuracy on a synthetic data set. The data set was generated using the dynamic model \eqref{eq:main_CO2} with some reasonable occupancy patterns. Figure \ref{fig:HMM_SYN} shows the true occupancy, the predicted occupancy (using a simple HMM model), and the simulated carbon dioxide concentration. We observe that the model prediction is systematically delayed after the occupancy change. This is a consequence of the fact that the simple HMM does not have any "memory". 

\begin{figure}[H]
    \captionsetup{justification=centering}
    \centering
    \includegraphics[scale=0.5]{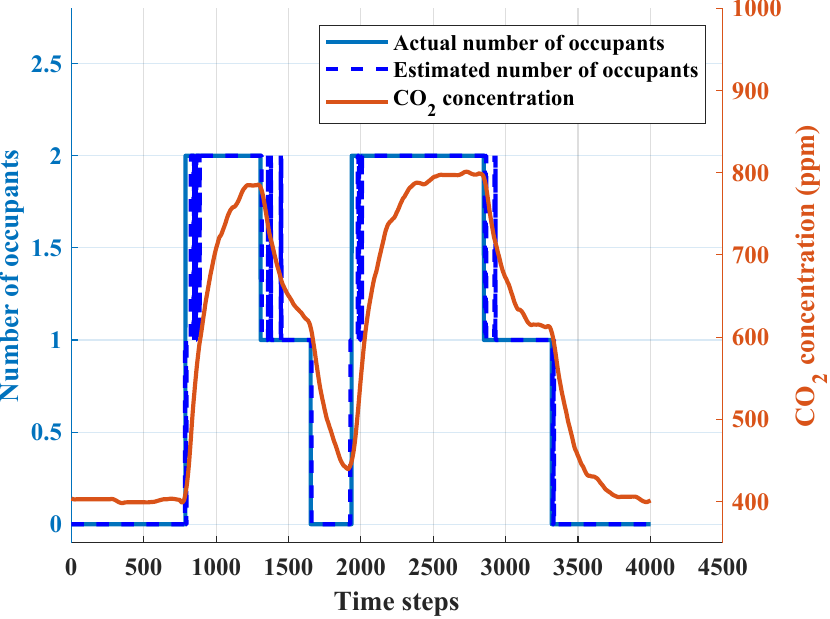}
    \caption{Performance of our proposed model on a synthetic dataset. The model is significantly more accurate than the simple HMM one, rapidly tracking most of the occupancy changes. The overall accuracy is a remarkable $97.26\%$. }
    \label{fig:ARHMM_SYN}
\end{figure}

Figure \ref{fig:ARHMM_SYN} shows the performance of our proposed model on the same synthetic dataset. Here, we see, that the model is able to quickly track the changes in occupancy (both increasing and decreasing). The model accuracy is rather impressive - $97.26\%$. For generating this synthetic data set we also made the ventilation rate to have two regimes (corresponding to different $\tau$) - $\tau_1=70$ and $\tau_2 = 100$ minutes, that are switched approximately once every few hours. Since our proposed model infers the auto-regressive coefficient, which is only a function of the ventilation rate, the model was able to very carefully track both occupancy and ventilation regime change.

\begin{figure}[H]
    \captionsetup{justification=centering}
    \centering
    \includegraphics[scale=0.5]{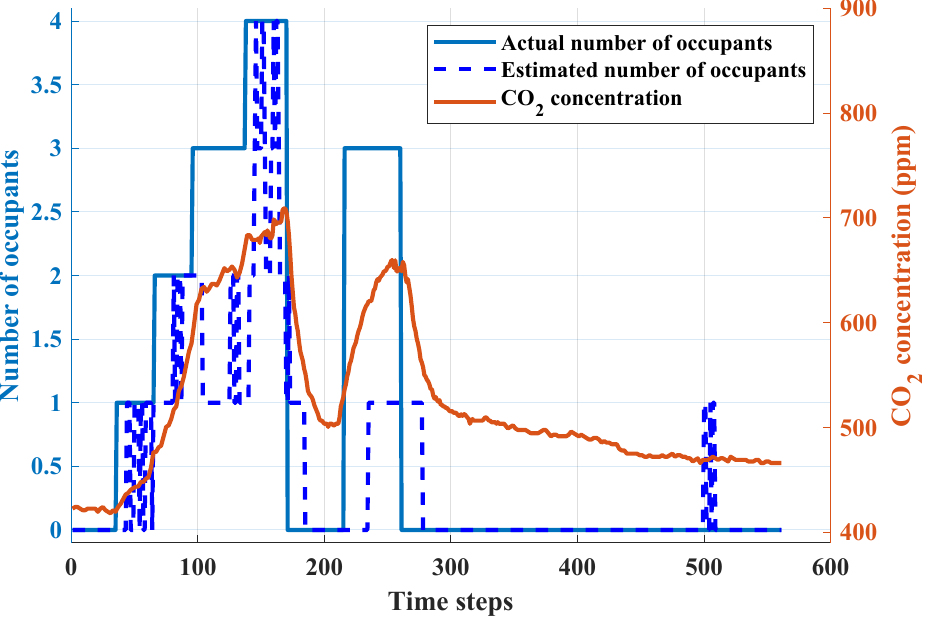}
    \caption{Performance of the simple HMM model on a real-life test-case. The model is distinctly delaying it's prediction and also tends to oscillate between the states frequently.  The overall accuracy is around $67.38\%$.}
    \label{fig:HMM_TS}
\end{figure}

Let us now turn to analyzing the real-life data. For this, we have performed an experiment with different numbers of occupants in one of the small seminar rooms at the EEE School at Nanyang Technological University. The room is approximately $3.5$ meters long and $2.5$ meters wide. We chose to change the occupancy approximately every 30 minutes with times of the arriving/departing people carefully tracked. Figure \ref{fig:HMM_TS} shows the performance of the simple HMM model for this test-case. Again, we notice that the model has systematic delays in tracking the room's occupancy. Moreover, it tends to oscillate between states too often. The overall model accuracy is $67.38\%$. 

\begin{figure}[H]
    \captionsetup{justification=centering}
    \centering
    \includegraphics[scale=0.5]{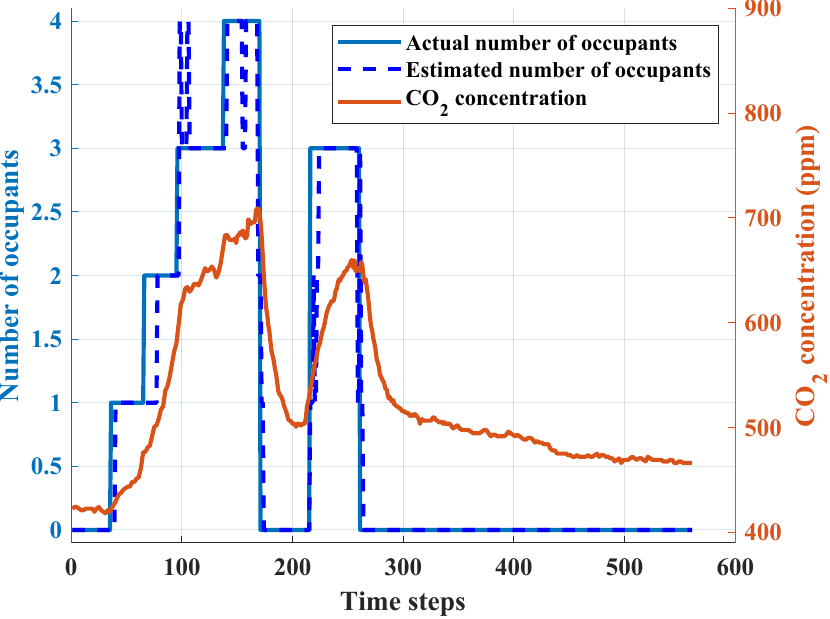}
    \caption{Performance of our proposed model on a real-life test-case. The model is remarkably accurate in tracking most changes in the occupancy. The overall accuracy is around $94.7\%$.}
    \label{fig:ARHMM_TS}
\end{figure}

Figure \ref{fig:ARHMM_TS} shows the performance of our proposed model for the same real-life test-case. We note that the model is very accurate and timely tracks the occupancy change most of the time. The overall performance is $94.7\%$.

\section{Conclusion and Further Research}

In this paper we have presented a new approach for building occupancy estimation by using a special type of switching process with Markov regime. Our model is derived from the underlying physical model of the carbon dioxide concentration dynamics, thus representing a case of \emph{physics-informed} learning. We have provided the physical interpretation of the model parameters and tested our model on two data sets - synthetic (simulated) and real-life. The model shows significant improvement compared to the simple HMM approach, consistently showing prediction accuracy above $90\%$. The model allows for different types of generalizations/improvements, some of which we will briefly summarize here. 

\begin{enumerate}
    \item Generalization of the model to multiple rooms with connected ventilation systems. This will provide the model with much more power for predicting occupancy in building under realistic conditions. The generalization can be done by representing the hidden states of every room as well as model coefficients as vectors. In this case, with the increase of the overall number of the model parameters, our physics-based approach will become even more valuable, allowing to choose the starting values of the model parameters close to their real values, thus avoiding different issues with algorithm performance (like overfitting). Also, the model can be generalized to process the data with incomplete observation of the building rooms. 

    \item Transforming the model to the online learning algorithm. Currently, the model analyzes the observation sequence and then fits the best possible sequence of the hidden states. We believe that it is possible to improve the model and make perform the assessment in real time, for example, using some sliding window of measurements and updating every measurement cycle (or few cycles). It is worth noting, that optimizing the model;s computation efficiency is an important task, since the model can then be used on-board of small-scale sensor devices where the computational power is limited. 

    \item The model can be extended by adding more observables, most likely temperature and humidity. Since these predictors can be highly correlated, certain modifications to the model algorithms should be done to avoid possible loss of performance. 
\end{enumerate}

\bibliography{bib/mainBib.bib}

\begin{thebibliography}{10}
\providecommand{\url}[1]{#1}
\csname url@samestyle\endcsname
\providecommand{\newblock}{\relax}
\providecommand{\bibinfo}[2]{#2}
\providecommand{\BIBentrySTDinterwordspacing}{\spaceskip=0pt\relax}
\providecommand{\BIBentryALTinterwordstretchfactor}{4}
\providecommand{\BIBentryALTinterwordspacing}{\spaceskip=\fontdimen2\font plus
\BIBentryALTinterwordstretchfactor\fontdimen3\font minus \fontdimen4\font\relax}
\providecommand{\BIBforeignlanguage}[2]{{%
\expandafter\ifx\csname l@#1\endcsname\relax
\typeout{** WARNING: IEEEtran.bst: No hyphenation pattern has been}%
\typeout{** loaded for the language `#1'. Using the pattern for}%
\typeout{** the default language instead.}%
\else
\language=\csname l@#1\endcsname
\fi
#2}}
\providecommand{\BIBdecl}{\relax}
\BIBdecl

\bibitem{IEA_efficiency}
\BIBentryALTinterwordspacing
IEA, ``Energy efficiency 2023,'' 2023. [Online]. Available: \url{https://www.iea.org/reports/energy-efficiency-2023}
\BIBentrySTDinterwordspacing

\bibitem{economidou2020review}
M.~Economidou, V.~Todeschi, P.~Bertoldi, D.~D'Agostino, P.~Zangheri, and L.~Castellazzi, ``Review of 50 years of eu energy efficiency policies for buildings,'' \emph{Energy and buildings}, vol. 225, p. 110322, 2020.

\bibitem{koch2011modeling}
S.~Koch, J.~L. Mathieu, D.~S. Callaway \emph{et~al.}, ``Modeling and control of aggregated heterogeneous thermostatically controlled loads for ancillary services,'' in \emph{Proc. PSCC}.\hskip 1em plus 0.5em minus 0.4em\relax Citeseer, 2011, pp. 1--7.

\bibitem{guo2021review}
M.~Guo, P.~Xu, T.~Xiao, R.~He, M.~Dai, and S.~L. Miller, ``Review and comparison of hvac operation guidelines in different countries during the covid-19 pandemic,'' \emph{Building and Environment}, vol. 187, p. 107368, 2021.

\bibitem{ding2020hvac}
J.~Ding, C.~W. Yu, and S.-J. Cao, ``Hvac systems for environmental control to minimize the covid-19 infection,'' \emph{Indoor and Built Environment}, vol.~29, no.~9, pp. 1195--1201, 2020.

\bibitem{allen2016associations}
J.~G. Allen, P.~MacNaughton, U.~Satish, S.~Santanam, J.~Vallarino, and J.~D. Spengler, ``Associations of cognitive function scores with carbon dioxide, ventilation, and volatile organic compound exposures in office workers: a controlled exposure study of green and conventional office environments,'' \emph{Environmental health perspectives}, vol. 124, no.~6, pp. 805--812, 2016.

\bibitem{chen2018building}
Z.~Chen, C.~Jiang, and L.~Xie, ``Building occupancy estimation and detection: A review,'' \emph{Energy and Buildings}, vol. 169, pp. 260--270, 2018.

\bibitem{rueda2020comprehensive}
L.~Rueda, K.~Agbossou, A.~Cardenas, N.~Henao, and S.~Kelouwani, ``A comprehensive review of approaches to building occupancy detection,'' \emph{Building and Environment}, vol. 180, p. 106966, 2020.

\bibitem{candanedo2017methodology}
L.~M. Candanedo, V.~Feldheim, and D.~Deramaix, ``A methodology based on hidden markov models for occupancy detection and a case study in a low energy residential building,'' \emph{Energy and Buildings}, vol. 148, pp. 327--341, 2017.

\bibitem{ai2014occupancy}
B.~Ai, Z.~Fan, and R.~X. Gao, ``Occupancy estimation for smart buildings by an auto-regressive hidden markov model,'' in \emph{2014 American Control Conference}.\hskip 1em plus 0.5em minus 0.4em\relax IEEE, 2014, pp. 2234--2239.

\bibitem{ephraim2002hidden}
Y.~Ephraim and N.~Merhav, ``Hidden markov processes,'' \emph{IEEE Transactions on information theory}, vol.~48, no.~6, pp. 1518--1569, 2002.

\bibitem{rabiner1989tutorial}
L.~R. Rabiner, ``A tutorial on hidden markov models and selected applications in speech recognition,'' \emph{Proceedings of the IEEE}, vol.~77, no.~2, pp. 257--286, 1989.

\bibitem{cali2015co2}
D.~Cal{\`\i}, P.~Matthes, K.~Huchtemann, R.~Streblow, and D.~M{\"u}ller, ``Co2 based occupancy detection algorithm: Experimental analysis and validation for office and residential buildings,'' \emph{Building and Environment}, vol.~86, pp. 39--49, 2015.

\bibitem{shumway2000time}
R.~H. Shumway, D.~S. Stoffer, and D.~S. Stoffer, \emph{Time series analysis and its applications}.\hskip 1em plus 0.5em minus 0.4em\relax Springer, 2000, vol.~3.

\bibitem{do2008expectation}
C.~B. Do and S.~Batzoglou, ``What is the expectation maximization algorithm?'' \emph{Nature biotechnology}, vol.~26, no.~8, pp. 897--899, 2008.

\bibitem{nguyen2005expectation}
H.~Nguyen and B.~C. Levy, ``The expectation-maximization viterbi algorithm for blind adaptive channel equalization,'' \emph{IEEE transactions on communications}, vol.~53, no.~10, pp. 1671--1678, 2005.

\bibitem{viterbi1967error}
A.~Viterbi, ``Error bounds for convolutional codes and an asymptotically optimum decoding algorithm,'' \emph{IEEE transactions on Information Theory}, vol.~13, no.~2, pp. 260--269, 1967.

\bibitem{churbanov2008implementing}
A.~Churbanov and S.~Winters-Hilt, ``Implementing em and viterbi algorithms for hidden markov model in linear memory,'' \emph{BMC bioinformatics}, vol.~9, pp. 1--15, 2008.

\end{thebibliography}

\end{document}